# Experimental demonstration of a μ=-1 metamaterial lens for magnetic resonance imaging

Manuel J. Freire, Ricardo Marques and Lukas Jelinek

*Department of Electronics and Electromagnetism, University of Seville*

*Facultad de Fisica, Avda. Reina Mercedes s/n, 41012 Seville, Spain*

*E-mail: freire@us.es, marques@us.es*

**In this work a μ=-1 metamaterial (MM) lens for magnetic resonance imaging (MRI) is demonstrated. MRI uses surface coils to detect the radiofrequency (RF) energy absorbed and emitted by the nuclear spins in the imaged object. The proposed MM lens manipulates the RF field detected by these surface coils, so that the coil sensitivity and spatial localization is substantially improved. Beyond this specific application, we feel that the reported results are the experimental confirmation of a new concept for the manipulation of RF field in MRI, which paves the way to many other interesting applications.**

After the demonstration of the ability of a slab of an ideal negative refractive index metamaterial (MM) with ε=-1 and μ=-1 to obtain sub-diffraction images[1], the issue of optical sub-wavelength imaging through the direct manipulation of the electromagnetic field has attracted a lot of attention. This effect has been shown in the optical frequency range[2], in the microwave range[3,4] and in the radiofrecuency range[5,6] by using different devices. However, the ability to image objects smaller than the wavelength is not a recent concept. It is something well known since long in magnetic resonance imaging (MRI), where imaged objects are very small as compared to the wavelength of the radiofrequency (RF) fields used to obtain the image. As it is well known, the generation of images in MRI is based on the detection of spatial variations in the phase and frequency of the RF energy absorbed and emitted by the nuclear spins of the imaged object[7]. These spatial variations are induced by some static magnetic field gradients, and the image is built from signals measured by a receiving coil that has not information about the relative location of the emitting magnetic dipoles. Conventional MRI involves many repeated measurements and then signal processing (inverse Fourier transforming) before obtaining an image of a single slice of tissue. Therefore, conventional sub-wavelength MRI is based on signal processing and does not involve any optical means such as focusing or collimation. At this point a question arises in a natural way: would it be possible to combine both signal processing and optical means, so that the ability of MM devices to directly obtain sub-wavelength images could be used to improve conventional MRI? Actually this issue was already explored to some extension by M. C. K. Wiltshire[8]. In his work, a microstructured MM with high magnetic permeability was used as a magnetic flux guide for MRI applications. By using this artificial medium, it was possible to put the receiving coil far from the imaged object and still to detect the image. However, it is not possible, by using this technique, to improve the quality of the images obtained with the coil directly placed on the skin of the patient. Anyway, this work clearly showed the compatibility of microstructured MMs with conventional MRI machines, as well as their potential usefulness in the frame of this technology, thus encouraging the search for further applications.

In a recent work[9] some of the authors proposed to use a sub-diffraction MM lens to improve the images obtained by surface coils in MRI. Surface MRI coils are usually placed just on the skin of the patient and are used to obtain images of tissues in the proximity of the coil. Due to its higher sensitivity, surface coils provide a signal-to-noise ratio (SNR) much larger than that obtained with whole-volume coils or body coils. However, whereas the sensitivity of body coils is uniform, the sensitivity of surface coils, as well as the SNR, decreases rapidly with the distance from the coil. Due to Lorentz reciprocity, the sensitivity of a coil is directly proportional to the intensity of the magnetic field created by the coil inside the body of the patient for a standard value of the current on the coil[10]. Fig. 1.a shows a typical plot of the sensitivity (i.e. the magnetic field intensity) of a circular coil placed on the skin of the patient. As it can be seen, this sensitivity decays with the distance from the coil, making the coil useless for obtaining images of tissues located at distances into the body deeper than the coil dimensions, typically. Let us now imagine that an ideal MM lens of thickness d is placed between the coil and the skin of the patient. An ideal MM lens of thickness d has the ability of focusing the electromagnetic field, translating the field distribution on the plane of the coil to another plane at a distance 2d from the coil, and vice-versa[1]. Therefore, this configuration would increase the coil sensitivity making it possible to obtain images of deeper tissues. Before to go further with the design, we must consider that RF fields used for MRI have an associated wavelength much higher than the dimensions of any practical coil. Therefore, we are in the realm of the quasi-magnetostatics, and a slab of a microstructured MM with µ=-1 should be enough for manufacturing the lens. Fig. 1.b shows the sensitivity of a coil in the presence of such a lens with µ=-1. The permeability of this MM lens was computed according to Eq. 13 in a previous work of the authors[11] and for the dimensions of the lens manufactured for the present work, which will be described in detail in next paragraphs. As it can be seen, the theory predicts a substantial increase of the coil sensitivity and SNR inside the body of the patient.

For the practical implementation of the aforementioned ideas, the ideal μ=-1 lens was mimicked by a slab consisting of a three-dimensional (3D) array of copper metallic rings loaded with non-magnetic capacitors. Capacitively-loaded rings (CLRs) were previously proposed by S. A. Schelkunoff[12] in order to design artificial media with strong magnetic response. In our case they were placed in a simple cubic lattice in order to obtain an isotropic artificial medium with μ=-1. The magnetic permeability of this medium was computed[11] as a function of the periodicity, the ring resistance, the ring self-inductance and the frequency of resonance. The fabricated MM lens was a two unit cells thick slab of this artificial medium. A sketch of the proposed lens is shown in Fig. 2.a. Before to proceed with the description of the experiments, some additional words will be devoted to the modelling of the lens. It is apparent that a slab made of only two layers of unit cells can hardly be considered as a continuous medium. Therefore, the detailed description of this structure deserves a deeper discussion. Actually, some of the authors have recently developed a homogenization procedure for thin slabs made of resonant metallic rings[13]. A conclusion of this analysis was that, for the specific configuration proposed in the present work, the continuous medium approach[11] gives a good description of the behaviour of the lens. Only a small shift in the frequency of operation of the lens with regard to that predicted by the homogenization procedure[13] was detected. This conclusion was confirmed by additional electromagnetic simulations made using the commercial electromagnetic solver *CST Microwave Studio.* Additional design corrections were necessary as a consequence of the finite size of the capacitors, which was not taken into account by the models. Finally, the lens was manufactured for operation in a MRI system of 1.5 Tesla (i.e., for a frequency of operation of 63.85 MHz). Fig. 2.b shows a sketch of a CLR of the lens, whose dimensions and design parameters are: external radius of the rings r=6.02 mm, ring width w=2.17 mm, ring self-inductance 13.45 nH, and ring capacitance 470 pF with a tolerance of 1%, which gives a resonance frequency of 63.28 MHz and a quality factor of 115. Finally, Fig. 2.c shows two photographs of the final device consisting of a 3D array of 18x18x2 cubic cells, with a periodicity of 15 mm and a total number of CLRs of 2196. The dimensions of the lens are 27x27x3 cm$^3$. The non-magnetic capacitors were provided by the company *American Technical Ceramics Corp.* (NY, USA). The rings were photoetched on FR4 substrate by the company *Circuitronica S. L.* (Seville, Spain) and the capacitors inserted by the company *Silicium S. L.* (Seville, Spain).

The fabricated lens was tested in a *General Electric* Signa 1.5T MRI machine using a standard 3-inch surface coil. In the experiment, one of the authors was lying on the MRI machine and the coil was placed besides one of his knees. In our study, axial images (i.e., images of a plane normal to the bore of the magnet) of type T1 were acquired using a standard spin-echo sequence typical of T1 acquisitions. The repetition time between signals or TR was 220 ms and the echo time (TE) was 10 ms. The field of view or FOV was 34 cm by 34 cm with a 256x192 data matrix. Two acquisitions with averaging were used in all cases. Fig. 3.a shows an axial image of the knees without the lens, so that both knees are touching. In this figure, the knee on the right of the image is closer to the coil and is clearly visible whereas the knee on the left is hardly visible, as it is expected from the fact that the sensitivity of the coil drops off rapidly with the distance. Fig. 3.b shows also an axial image of both knees with the lens being placed between them. In spite of the fact that the distance between the coil and the knee on the left of the image is larger than in the absence of the lens, this knee is now more visible due to the presence of the lens. This makes apparent that the lens increases the sensitivity of the coil. We feel that this is a completely new result that relies on the specific design of the lens, which mimics a µ=-1 medium. We also feel that this results introduces a new concept in MRI, showing that a µ=-1 MM lens can be applied in the frame of conventional MRI technology in order to improve the sensitivity of surface coils.

Up to now we have shown that MM lenses can be useful in MRI technology due to its ability of focusing the RF magnetic field lines of force. Specifically, we have shown that MM lenses can be applied to improve the sensitivity of surface coils, thus resulting in an improvement of image quality, reduction of acquisition time and/or increase of spatial localization. Since MM lenses can translate the field distribution in a plane behind the lens to another equivalent plane in front of the lens, they can be also useful for obtaining images of deeper tissues. We feel that the reported results provide a sufficient proof of concept for the reported effect. Other MM lens configurations[14] different from the manufactured lens could be also useful for this application. Regarding further applications of this new concept, we feel that it may also find application in parallel imaging MRI technology[15], as suggested previously by the authors[14]. MRI parallel imaging techniques use several surface coils and take advantage of the spatial localization of the images detected by each coil[16] to reduce acquisition time. The spatial localization of the images detected by the different coils would be substantially increased if the coils were placed on the equivalent plane of the imaged slice of tissue. In the limit, this technique could make possible to avoid the phase encoding process[17], thus opening the way to real time image acquisition of deep tissues.

This work has been supported by the Spanish Ministerio de Educación y Ciencia under projects TEC2007-68013-C02-01/TCM and CSD2008-00066, and by Spanish Junta de Andalucía under project P06-TIC-01368. We want to thank Dr. Francisco Moya, Dr. Eduardo Gil and Borja Mohedano, from PET Cartuja Medical Center (Seville), for providing the MRI facilities used in this work and for their advice.

Figure 1: Normalized calculated sensitivity for different distances in cm of (a) a standard surface coil of three inch of diameter and (B) of the same coil placed on the lens. Units are arbitrary.

Figure 2: (a) Sketch of the lens: a 3D array of capacitively loaded rings (CLRs). (b) Sketch of a CLR with dimensions. Parameters of the fabricated CLRs: w=2.17 mm, r=6.02 mm, self-inductance 13.45 nH, capacitance 470 pF with tolerance of 1%, resonance frequency 63.28 MHz, quality factor Q=115. (c) Photographs of the fabricated lens consisting of a 3D array of 18x18x2 cubic cells with a periodicity of 15 mm. The total number of CLRs is 2196. The dimensions of the lens are: 27x27x3 cm$^3$.

Figure 3: Axial T1 image of the knees of one of the authors (a) without the lens, and (b) with the lens between the knees. It must be noted that the magnetic resonance images are inverted with respect to the photographs, and that this is inherent to the MRI acquisition process.

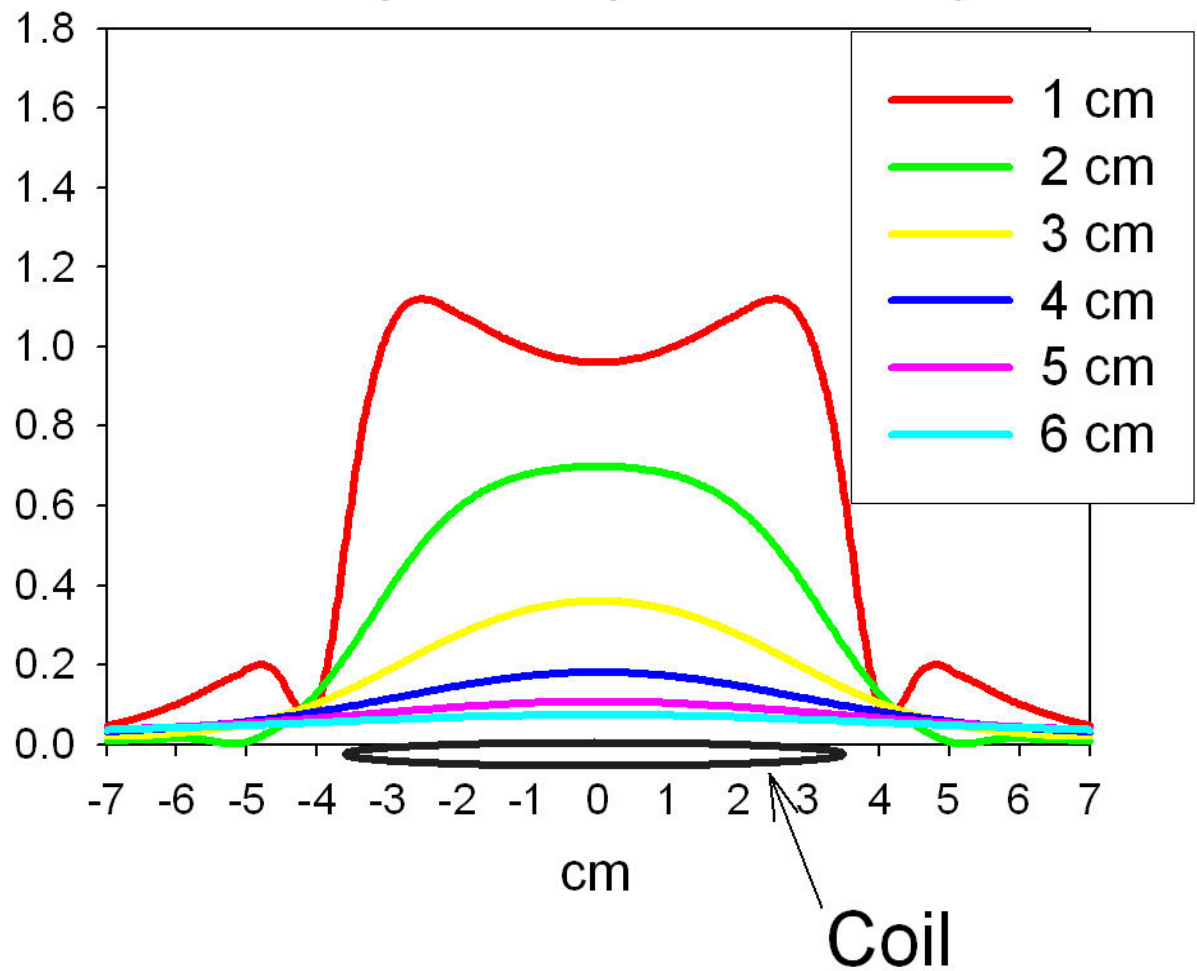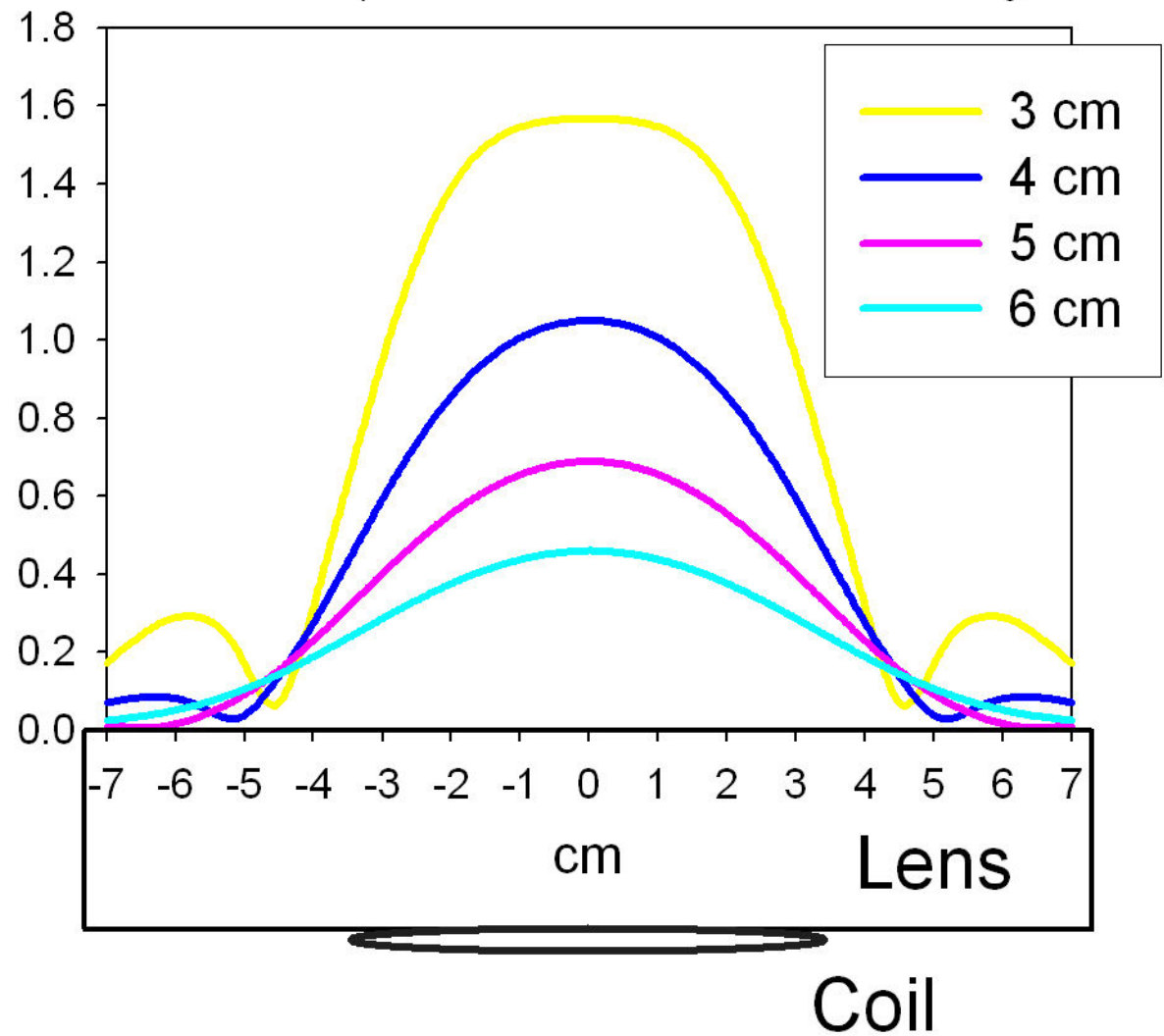

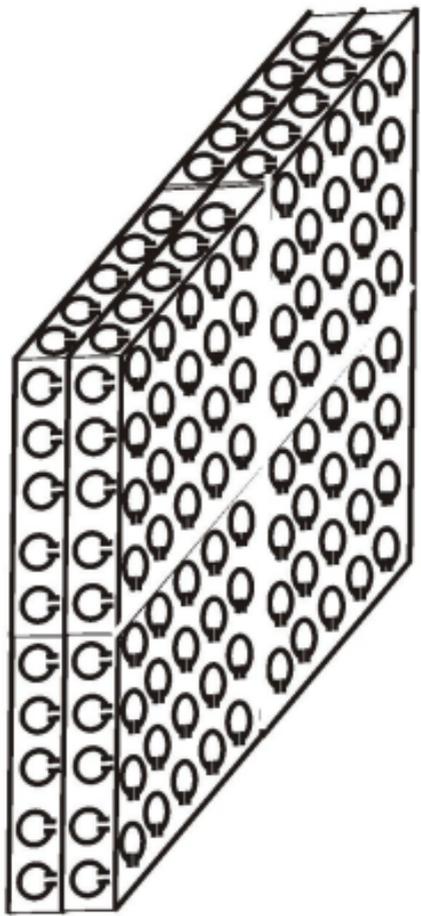 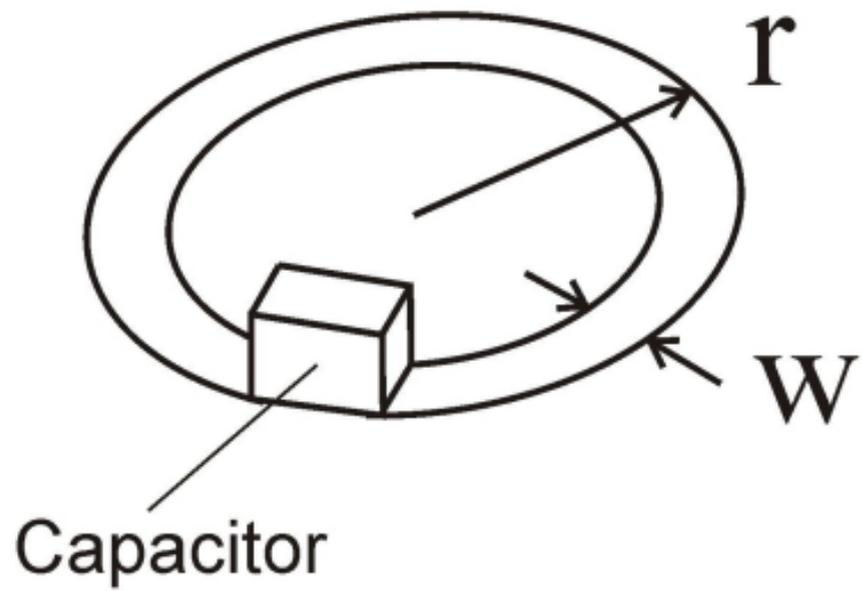 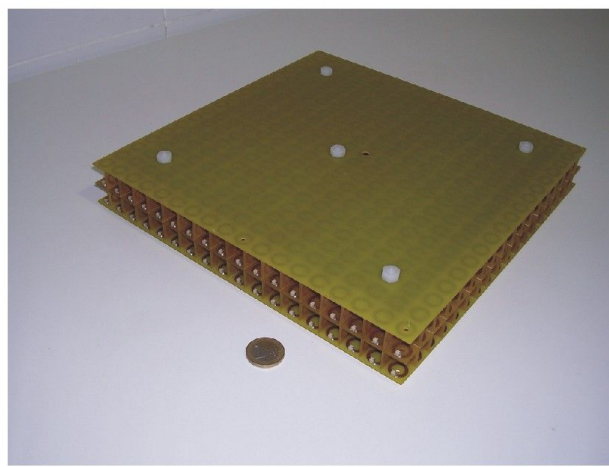 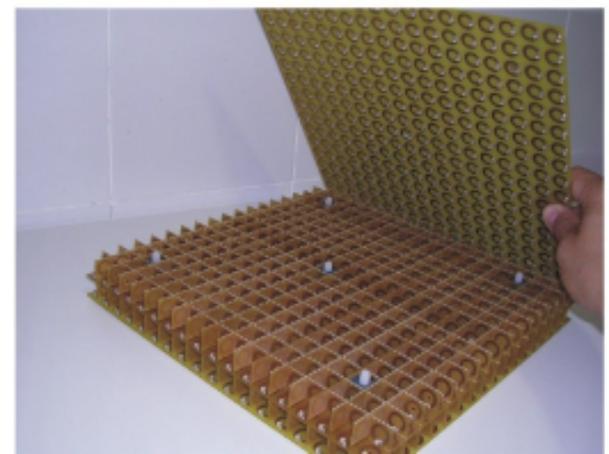

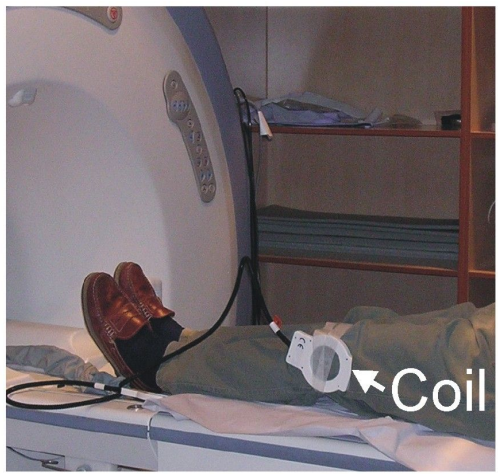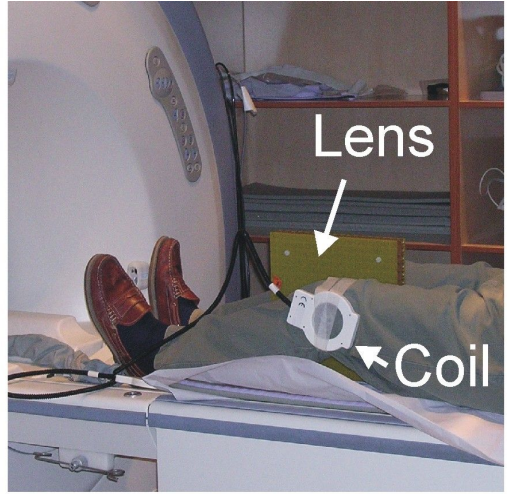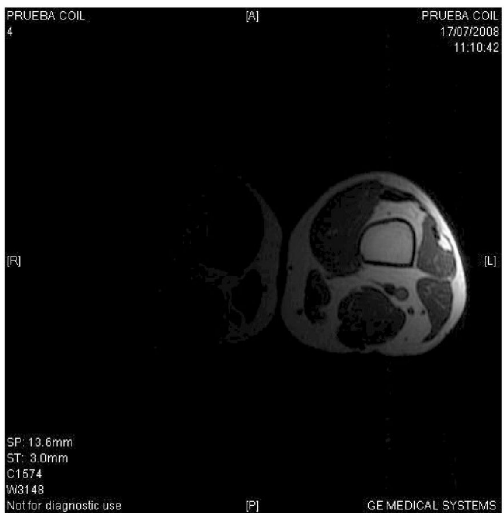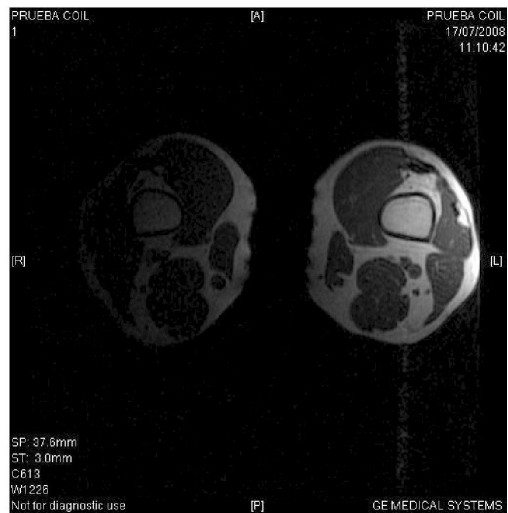